\documentclass[aps,prb,showpacs,twocolumn,superscriptaddress]{revtex4-1}
\usepackage{graphicx,amsmath,amssymb}
\usepackage[usenames]{color}
\usepackage{indentfirst}
\usepackage{float}
\usepackage[T1]{fontenc}
\usepackage[colorlinks=true, citecolor=blue, urlcolor=blue, linkcolor=blue ]{hyperref}
\hypersetup{breaklinks=true}
\begin{document}

\bibliographystyle{apsrev4-1}

\title{Hidden Anderson Localization in Disorder-Free Ising-Kondo Lattice}
\author{Wei-Wei Yang, Lan Zhang, Xue-Ming Guo, Yin Zhong}
\email{zhongy@lzu.edu.cn}
\affiliation{School of Physical Science and Technology $\&$ Key Laboratory for
Magnetism and Magnetic Materials of the MoE, Lanzhou University, Lanzhou 730000, China}

\begin{abstract}
Anderson localization (AL) phenomena usually exists in systems with random potential. However, disorder-free quantum many-body systems with local conservation can also exhibit AL or even many-body localization transition. In this work, we show that the AL phase exists in a modified Kondo lattice without external random potential. The density of state, inverse participation ratio and temperature-dependent resistance are computed by classical Monte Carlo simulation, which uncovers the AL phase from previously studied Fermi liquid and Mott insulator regime. The occurrence of AL roots from quenched disorder formed by conservative localized moments. Interestingly, a many-body wavefunction is found, which captures elements in all three paramagnetic phases and is used to compute their quantum entanglement. In light of these findings, we expect the disorder-free AL phenomena can exist in generic translation-invariant quantum many-body systems.
\end{abstract}

\maketitle
\section{Introduction}
Localization phenomena due to random potential, namely the Anderson localization (AL), is at the heart position in modern condensed matter physics\cite{Anderson,Lee,Evers}. When including the effect of interaction, many-body localization (MBL) emerges and has spurred intensive studies on disordered quantum many-body systems\cite{Gornyi,Basko,Nandkishore,Abanin}.
Interestingly, if local conservation, e.g. $Z_{2}$ gauge symmetry, exists in Hamiltonian, AL/MBL can exist without external quenched disorder, thus certain translation-invariant quantum systems can exhibit AL or MBL\cite{Antipov,Castro,Smith,Smith2}. However, existing examples of disorder-free AL and MBL are still rare, in spite of general interests on their relation to quantum thermalization, lattice gauge field, topological order and novel quantum liquid\cite{Srednicki,Rigol,Kogut,Smith2,Castro,Smith3}.

\begin{figure}
\includegraphics[width=1.2\linewidth]{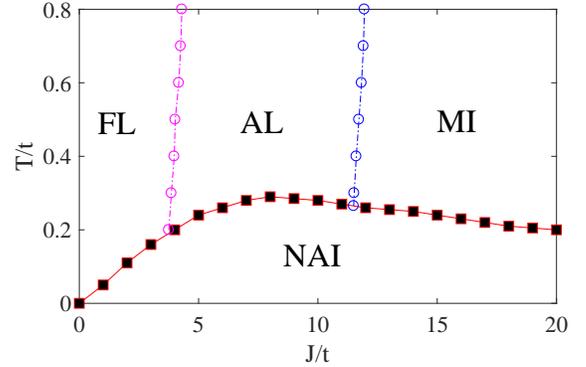}
\caption{\label{fig:1} Finite temperature phase diagram of Ising-Kondo lattice (IKL) model on square lattice (Eq.~\ref{eq1}) from classical Monte Carlo (MC) simulation. There exist Fermi liquid (FL), Mott insulator (MI), N\'{e}el antiferromagnetic insulator (NAI) and an Anderson localization (AL) phase. }
\end{figure}

Recently, we have revisited a modified Kondo lattice model, namely the Ising-Kondo lattice (IKL)\cite{Sikkema}, which is shown to reduce to fermions moving on static potential problem due to local conservation of localized moments, thus admits a solution by classical Monte Carlo (MC) simulation\cite{Yang}. On square lattice at half-filling, Fermi liquid (FL), Mott insulator (MI) and N\'{e}el antiferromagnetic insulator (NAI) are established. (See Fig.~\ref{fig:1}.) When doping is introduced, spin-stripe physics emerges with competing magnetic ordered states, similar to $t-J$ model and $f$-electron materials\cite{White,Lynn}. As emphasized by Antipov et. al. in the context of Falicov-Kimball model\cite{Antipov,Falicov}, since the weight of static potential satisfies Boltzmann distribution, and if temperature is high enough, the probability distribution of all configurations of static potential tends to be equal, thus may realize binary random potential distribution. When such intrinsic random potential is active, AL of fermions appears without external quenched disorder.

In this paper, we explore the possibility of accessing AL in IKL model on square lattice. To simplify the discussion and meet with our previous work, here we focus on half-filling case though doping the half-filled system does not involve any technical difficulty. (An example on doped system is given in Appendix~\ref{apD}.) By inspecting density of state (DOS), inverse participation ratio (IPR) and temperature-dependent resistance, we will show that the AL phase emerges in intermediate coupling regime between metallic FL and insulating MI above antiferromagnetic critical temperature. (See Fig.~\ref{fig:1}.) The occurrence of AL results from quenched disorder, formed by the conservative localized moments at each site. Interestingly, we find a many-body wavefunction, which captures elements in all three paramagnetic phases and is used to compute their entanglement entropy. In light of these findings, the disorder-free AL phenomena could exist in more generic translation-invariant quantum many-body systems.

The remainder of this paper is organized as follows: In Sec.~\ref{sec2}, the IKL model is introduced and explained. In Sec.~\ref{sec3}, MC simulation is performed and observables such as DOS, IPR and temperature-dependent resistance are computed. Analysis on MC data suggests the appearance of AL in intermediate coupling regime where thermal fluctuation destroys magnetic long-ranged order. A wavefunction is constructed and the entanglement entropy is evaluated. Sec.~\ref{sec4} gives a summary and a brief discussion on AL in generic quantum many-body systems.

\section{Model}\label{sec2}
The IKL model on square lattice at half-filling is defined as follows\cite{Yang}
\begin{eqnarray}
\hat{H}&&=-t\sum_{\langle i,j\rangle,\sigma}\hat{c}_{i\sigma}^{\dag}\hat{c}_{j\sigma}+\frac{J}{2}\sum_{j\sigma}\hat{S}_{j}^{z} \hat{c}_{j\sigma}^{\dag}\hat{\sigma^z}\hat{c}_{j\sigma}\label{eq1}
\end{eqnarray}
where itinerant electron interacts with localized $f$-electron moment via longitudinal Kondo exchange. Here, $\hat{\sigma}$ is the spin operator of conduction electron; $\hat{c}_{j\sigma}^\dagger$ is the creation operator of conduction electron;  $\hat{S}_{j}^{z}$ denotes the localized moment of $f$-electron at site $j$. $t$ is the hopping integral between nearest-neighbor sites $i,j$ and $J$ is the longitudinal Kondo coupling, which is usually chosen to be antiferromagnetic ($J>0$). In literature, this model (with $x$-axis anisotropy) is originally proposed to account for the anomalously small staggered magnetization
and large specific heat jump at hidden order transition in URu$_{2}$Si$_{2}$\cite{Sikkema,Mydosh}. It can explain
the easy-axis magnetic order and paramagnetic metal or bad metal behaviors in the global phase diagram of heavy fermion compounds\cite{Coleman2015,Si,Coleman}. We have to emphasize that due to the lack of transverse Kondo coupling, enhancement of effective mass and related Kondo screening observed in many heavy fermion compounds are not captured by IKL model.

In Ref.\onlinecite{Yang}, we have observed that $f$-electron's spin/localized moment at each site is conservative since $[\hat{S}_{j}^{z},\hat{H}]=0$. Therefore, taking the eigenstates of spin $\hat{S}_{j}^{z}$ as bases, the Hamiltonian Eq.~\ref{eq1} is automatically reduced to a free fermion moving on effective static potential $\{q_{j}\}$
\begin{eqnarray}
\hat{H}(q)=-t\sum_{\langle i,j\rangle,\sigma}\hat{c}_{i\sigma}^{\dag}\hat{c}_{j\sigma}+\sum_{j\sigma}\frac{J}{4}q_{j} \hat{c}_{j\sigma}^{\dag}\hat{\sigma^z}\hat{c}_{j\sigma}.\label{eq2}
\end{eqnarray}
 Here, $q$ emphasizes its $q$ dependence and $\hat{S}_{j}^{z}|q_{j}\rangle=\frac{q_{j}}{2}|q_{j}\rangle, q_{j}=\pm1$. Now, the many-body eigenstate of original model Eq.~\ref{eq1} can be constructed via single-particle state of effective Hamiltonian Eq.~\ref{eq2} under given configuration of effective Ising spin $\{q_{j}\}$. So, Eq.~\ref{eq1} is solvable in the spirit of well-known Kitaev's toric-code and honeycomb model\cite{Kitaev1,Kitaev2}. It can be regarded as the spinful version of the Falicov-Kimball model\cite{Antipov,Falicov}. At finite-$T$, this model can be readily simulated by classical MC simulation\cite{Czajka}. (We consider periodic $N_{s}=L\times L$ lattices with $L$ up to $20$. Details on MC can be found in Appendix~\ref{apA})

A careful reader may note that isotropic Kondo lattice with ferromagnetic coupling has been studied by classical MC simulation\cite{Yunoki,Dagotto}. The algorithm of MC in our model is similar to those models, while they consider the ferromagnetic coupling (we consider antiferromagnetic coupling), thus their resultant phase diagram is rather different from ours. Technically, in their model, the local spin is approximated with classical vector, then the MC simulation is done by sampling classical configurations of those vectors. In contrast, our model is the Ising version of Kondo lattice models, and the local spin is the quantum Ising spin. When we choose their eigenstates, such quantum spin is exactly transformed to classical spin and MC is done straightforwardly. Therefore, our model is exactly simulated by MC while the ferromagnetic isotropic Kondo lattice is approximately simulated.
\section{Result}\label{sec3}
In terms of MC, we have determined the finite temperature phase diagram in Fig.~\ref{fig:1}. In addition to well-established FL, MI and
NAI in previous work, interestingly, we find an AL phase in intermediate coupling regime at high $T$. There is no transition but crossover from FL to AL and AL to MI. Since the former three phases have been clearly studied\cite{Yang}, in this work, we focus on the AL phase.

To characterize the AL phase from FL or MI, we study the DOS, IPR and temperature-dependent resistance of conduction electrons\cite{Evers,Antipov}. The DOS of $c$-fermion $N(\omega)$ is evaluated from
\begin{equation}
\mathrm{N}(\omega)\simeq\frac{1}{N_{m}N_{s}}\sum_{\{q_{j}\}}\sum_{n\sigma}\delta(\omega-E_{n\sigma}(q))\label{eq6}.
\end{equation}
In FL, its DOS at Fermi energy ($N(0)$, $\omega=0$ is Fermi energy) is finite. For usual AL phase, it results from localization-delocalization transition from metallic FL states caused by random potential, so its $N(0)$ is finite. As for MI, Mott gap driven by local magnetic fluctuation leads to vanishing $N(0)$\cite{Yang}.

The IPR measures tendency of localization. Here the energy/frequency-dependent IPR is used,
\begin{equation}
\mathrm{IPR}(\omega)\simeq\frac{1}{N_{m}N_{s}}\sum_{\{q_{j}\}}\sum_{n\sigma}\sum_{j}\delta(\omega-E_{n\sigma}(q)) (\phi_{n\sigma}^{j})^{4}.\label{eq7}
\end{equation}
In a localized state, it has to saturate for large system size. In contrast, in a delocalized state, such as FL, it has size-dependence as $\mathrm{IPR}\propto 1/V$, suggesting a well-defined inverse-volume behavior\cite{Evers,Antipov}.

The temperature-dependent resistance is related to static conductance $\sigma_{dc}$ as $\rho=1/\sigma_{dc}$, which reads
\begin{eqnarray}
\sigma_{dc}=\frac{\pi t^{2}e^{2}}{N_{m}}\sum_{\{q_{j}\}}\int d\omega\frac{-\partial f_{F}(\omega)}{\partial \omega}\Phi^{q}(\omega).\label{eq8}
\end{eqnarray}
The derivation of $\sigma_{dc}$ and the detailed form of $\Phi^{q}(\omega)$ can be found in Appendix~\ref{apC}. Generally, localized phases show insulating behavior at low temperature while delocalized metallic phases have contrary tendency.
\begin{figure}[htb]
\includegraphics[width=1.1\linewidth]{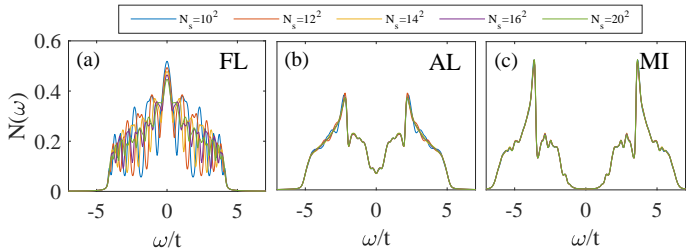}
\caption{\label{fig:2} Density of state (DOS) of conduction electron $N(\omega)$ in (a) FL ($J/t=2$), (b) AL ($J/t=8$) and (c) MI ($J/t=14$) phases at $T/t=0.4$.}
\end{figure}

\begin{figure}
\includegraphics[width=1.1\linewidth]{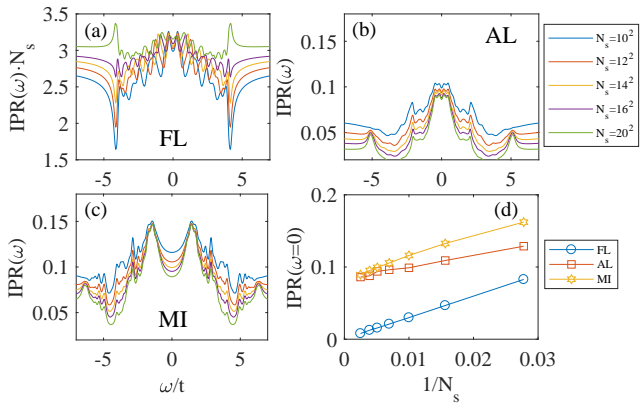}
\caption{\label{fig:3} Inverse participation ratio (IPR) of conduction electron $\mathrm{IPR}(\omega)$ in (a) FL ($J/t=2$), (b) AL ($J/t=8$) and (c) MI ($J/t=14$) phases at $T/t=0.4$. (d) shows finite-size extrapolation of IPR at Fermi energy $\omega=0$.}
\end{figure}

\begin{figure}
\includegraphics[width=1.0\linewidth]{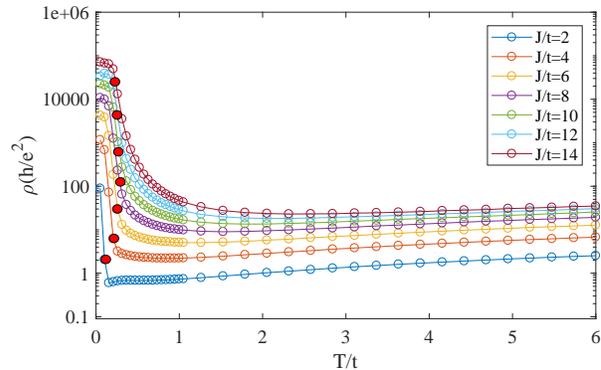}
\caption{\label{fig:4} Temperature-dependent resistance of conduction electron $\rho$ versus $T$ for different Kondo coupling $J/t$. Red dots indicate magnetic critical temperature $T_{c}$.}
\end{figure}

Now, from MC calculation of these quantities, e.g., Fig.~\ref{fig:2}, \ref{fig:3} and \ref{fig:4}, we find an AL phase in intermediate coupling, beside well-established FL and MI at high $T$ regime. In Fig.~\ref{fig:2}, AL has finite $N(0)$ though its strength is heavily suppressed and looks like a pseudogap. The reason is that due to the preformed local antiferromagnetic order, the band gap begins to form at low $T$. When increasing temperature, excitation of localized moments appears and it acts like impurity scattering center in the well-formed antiferromagnetic background. Then, the conduction electron scatters from such impurity and contributes impurity bound state, which fills in the band gap\cite{Thoss}. At high $T$, the long-ranged antiferromagnetic order melts but the band gap survives due to remaining local antiferromagnetic order. After considering impurity bound states, $N(0)$ in AL is finite and a pseudogap-like behavior appears.

In Fig.~\ref{fig:3}, we note that IPR of FL satisfies the expected inverse-volume law while AL and MI have saturated IPR around $\omega=0$.
Additionally, extrapolation of IPR at $\omega=0$ into infinite system size indicates that the localization length in FL is infinite while AL and MI only have finite localization length.

The $T$-dependent resistance of conduction electron is shown in Fig.~\ref{fig:4}, and the crossover from FL to AL and MI is clearly demonstrated. In both AL and MI, the insulating behaviors appear before the formation of insulating NAI, ($T>T_{c}$) which suggests they are insulators driven by correlation and thermal fluctuation.

\subsection{Why AL appears}
As found by MC simulation, the AL phase appears in intermediate coupling regime above the magnetic long-ranged ordered state.
If $T$ is high enough, the effective Boltzmann weight $\rho(q)$ for given configuration of static potential/Ising spin
should be equal. Thus, $c$-fermion feels an effective potential, which works as binary random potential. (Recall that $q_{j}=\pm1$ has two values.) Averaging over $\rho(q)$ leads to a disorder average for $c$-fermion and the AL phase is realized.

Technically, the above statement means that
\begin{equation}
\langle\hat{O}\rangle_{T\rightarrow\infty}\simeq\frac{1}{N_{m}}\sum_{\{q_{j}\}} (\hat{O}^{q}(q)+\langle\langle\hat{O}^{c}(q)\rangle\rangle),\label{eq9}
\end{equation}
where each configuration $\{q_{j}\}$ can be randomly chosen from all possible $2^{N_{s}}$ configurations rather than the ones weighted by $\rho(q)$. ($N_{m}$ is the number of configuration. For $N_{s}\sim10^{2}-10^{3}$, $N_{m}\sim10^{3}$ is used.) If $J/t\gg1$, one expects the appearance of AL phase. Considering that Mott localization due to correlation should dominate at strong coupling, we recover the finding that the AL phase occurs in intermediate coupling regime.

\begin{figure}
\includegraphics[width=1.1\linewidth]{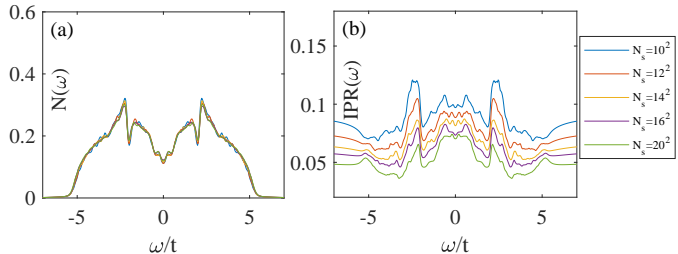}
\caption{\label{fig:5} The DOS and IPR for $J/t=8$ at effective temperature $T=\infty$.}
\end{figure}

To justify above argument, we show DOS and IPR in Fig.~\ref{fig:5} using Eq.~\ref{eq9},
where both DOS and IPR agree with ones in AL phase in Fig.~\ref{fig:2}~(b) and Fig.~\ref{fig:3}~(b) in intermediate coupling regime ($J/t=8$). Thus, it suggests again that AL phase in our model results from effective random potential formed by localized moment. As a matter of fact, if we consider weak and strong coupling case, their DOS and IPR are similar to the counterpart in Fig.~\ref{fig:2} and \ref{fig:3} as well, thus all three states at high-$T$ are stable in $T=\infty$ limit.

\subsection{A many-body wavefunction for all three paramagnetic states}
Motivated by arguments in last subsection and Eq.~\ref{eq9}, we write down the following many-body state, which approximates FL, AL and MI,
\begin{equation}
|\Psi\rangle=\frac{1}{\sqrt{N_{m}}}\sum_{\{q_{j}\}}|q_{1},q_{2},...,q_{N_{s}}\rangle\otimes|\psi\rangle\label{eq10}
\end{equation}
with $|\psi\rangle$ being the many-body eigenstate of effective free fermion Hamiltonian Eq.~\ref{eq2} for given configuration $\{q_{j}\}$, $J$ and the construction of configuration $\{q_{j}\}$ are identical to Eq.~\ref{eq9}. It is readily shown that $\langle \Psi|\hat{O}|\Psi\rangle$ gives the same result as Eq.~\ref{eq9}. Therefore, Eq.~\ref{eq10} itself behaves as a thermal statistical ensemble at $T=\infty$.

\subsection{Quenched disorder}

In the context of AL, both quenched disorder and annealed disorder are static disorder. In literature, the difference between these two kinds of disorder is that, by definition, quenched disorder corresponds to a time-independent probability distribution function (PDF) \cite{Antipov,PhysRevB.90.024202,PhysRevB.76.245122,RevModPhys.58.801,Byczuk}:
\begin{equation}
P(\{q_1,q_2,...,q_{N_s}\})=\prod_{i=1}^{N_s}P(q_i),
\label{eq10.2}
\end{equation}
where each of the $P(q_i)$ is the same. In contrast to the PDF in quenched disorder, the annealed disorder follows a thermal distribution.

In our model, the quenched disorder means that the configuration of local moments is randomly chosen at every site, where $P(q_i)$ cannot fluctuate in time. We have realized the quenched disorder at the infinite temperature ($T=\infty$) situation. A quenched disordered many-body wavefunction is constructed with $N_m$ randomly chosen configurations of local moments. Here each $P(q_i)$ is the same and time-independent, thus leads to a global constant-type $P(\{q_1,q_2,...,q_{N_s}\})$ as well. 
Under this constant-type PDF, the many-body wavefunction could be regarded as quenched disordered state \cite{Antipov}.

As mentioned above, with this many-body wavefunction we can cover the main results of MC simulation at high temperature. It suggests that the quenched disordered state do capture the main physics of IKL at high temperature region. Therefore, it is reasonable to attribute the existence of disorder-free AL to the intrinsic quenched disorder of local moments.

\subsection{Entanglement entropy}
The entanglement entropy $S_{EE}$ is used to characterize the universal quantum correlation in many-body state. For our model, we calculate $S_{EE}$ for each Slater determinant state $|\psi\rangle$ in given Ising configuration $\{q_{j}\}$ as follows\cite{Dzero}.

We consider open boundary condition and use $|\psi\rangle$ to compute equal-time correlation function $g_{ij,\sigma}^{q}$ as $g_{ij,\sigma}^{q}=\langle\psi|\hat{c}_{i\sigma}^{\dag}\hat{c}_{j\sigma}|\psi\rangle$. Dividing our system into $A$ and $B$ parts,
a reduced correlation function $\tilde{g}_{ij,\sigma}^{q}$ is constructed as
\begin{equation}
\tilde{g}_{ij,\sigma}^{q}=g_{ij,\sigma}^{q}~~~~ for~~i,j\in A\nonumber
\end{equation}
and others are zero. Treating $\tilde{g}_{\sigma}^{q}$ as $N_{A}\times N_{A}$ matrix with non-zero eigenvalues $\{\xi_{\alpha}\}$ and $N_{A}$ is the number of sites in region $A$. The entanglement entropy between subsystem $A,B$ is
\begin{equation}
S_{EE}=-\sum_{\alpha}\left[\xi_{\alpha}\ln\xi_{\alpha}+(1-\xi_{\alpha})\ln(1-\xi_{\alpha})\right].\label{eq11}
\end{equation}

\begin{figure}
\includegraphics[width=1.0\linewidth]{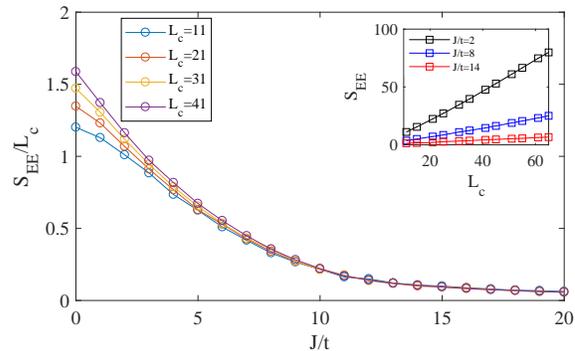}
\caption{\label{fig:6} The entanglement entropy $S_{EE}$ of many-body wavefunction Eq.~\ref{eq10} for different boundary size $L_{c}$ between two subsystems and different Kondo coupling $J$.}
\end{figure}

Using Eq.~\ref{eq11}, we have computed $S_{EE}$ for many-body state Eq.~\ref{eq10}, and the results are shown in Fig.~\ref{fig:6}.
Firstly, $S_{EE}$ decreases monotonically from samll-$J$ to large-$J$ regime, agreeing with the increased localization tendency. Another interesting feature is that when $J/t\simeq12$, $S_{EE}/L_{c}$ collapses into a single line, thus indicating a crossover from AL to MI at $T=\infty$. More close inspection on $S_{EE}$ shows a linear-dependence on $L_{c}$ in MI regime, which is the well-known area-law for $S_{EE}$\cite{Eisert}. For FL regime, its $S_{EE}$ deviates from area-law with a logarithmic correction\cite{Wolf,Gioev}. A fitting in FL gives $S_{EE}\simeq 0.275L_{c}\ln L_{c}+5.05$ for $J/t=2$. As for AL, it qualitatively obeys area-law though a small deviation exists due to inevitable mixing with delocalized states.

\section{Conclusion and Discussion}\label{sec4}
In conclusion, we have established an AL phase in a modified Kondo lattice model without external disorder potential. The presence of AL results from quenched disorder, formed by conservative localized moment at each site and is a stable phase even at infinite temperature. A many-body wavefunction is constructed to understand AL, FL and MI. Their entanglement entropy is computed and the area-law is violated in FL. In light of these findings, we recall that interacting many-electron system can be rewritten as free electrons moving on fluctuated background field after Hubbard-Stratonovich transformation\cite{Hubbard,Stratonovich}. As noted by Antipov et. al.\cite{Antipov}, if dynamics of the background field is frozen, it can act as a random potential to scatter electrons. Then, AL or even MBL phases could be observed in generic many-body Hamiltonian. It will be interesting to see whether the isotropic Kondo lattice (transverse Kondo coupling is included), which is the standard model for heavy fermion study, will support the presence of those localized states of matter.

\section*{Acknowledgments}
This research was supported in part by NSFC under Grant No.~$11704166$, No.~$11834005$, No.~$11874188$.

\appendix
\section{MC simulation}\label{apA}
With MC simulation, we can write the partition function as
\begin{equation}
\mathcal{Z}=\mathrm{Tr}e^{-\beta \hat{H}}=\mathrm{Tr}_{c}\mathrm{Tr}_{S}e^{-\beta \hat{H}}=\sum_{\{q_{j}\}}\mathrm{Tr}_{c}e^{-\beta \hat{H}(q)}.\nonumber
\end{equation}
Here, the trace is split into $c$-fermion and $\hat{S}^{z}$, where the latter is transformed into the summation over all possible configuration $\{q_{j}\}$. For each single-particle Hamiltonian $\hat{H}(q)$, it can be easily diagonalized into
\begin{equation}
\hat{H}(q)=\sum_{n\sigma}E_{n\sigma}\hat{d}_{n\sigma}^{\dag}\hat{d}_{n\sigma}\nonumber
\end{equation}
where $E_{n\sigma}$ is the single-particle energy level and $\hat{d}_{n\sigma}$ is the quasi-particle. The fermion $\hat{d}_{n\sigma}$ is related into $\hat{c}_{j\sigma}$ via
\begin{eqnarray}
\hat{c}_{j\sigma}&&=|0\rangle\langle j\sigma|=\sum_{n}|0\rangle\langle j\sigma|n\sigma\rangle\langle n\sigma|\nonumber\\
&&=\sum_{n}|0\rangle\langle n\sigma|\langle j\sigma|n\sigma\rangle=\sum_{n}\hat{d}_{n\sigma}\phi_{n\sigma}^{j}\nonumber.
\end{eqnarray}
with $\phi_{n\sigma}^{j}\equiv\langle j\sigma|n\sigma\rangle$. Now, the trace over $c$-fermion can be obtained as
\begin{eqnarray}
\mathrm{Tr}_{c}e^{-\beta \hat{H}(q)}&&=\sum_{n\sigma}\langle n\sigma|e^{-\beta \sum_{m\sigma'}E_{m\sigma'}\hat{d}_{m\sigma'}^{\dag}\hat{d}_{m\sigma'}}|n\sigma\rangle\nonumber\\
&&=\prod_{n\sigma}(1+e^{-\beta E_{n\sigma}}).\nonumber
\end{eqnarray}
This is the familiar result for free fermion, however one should keep in mind that $E_{n\sigma}$ actually depends on the effective Ising spin configuration $\{q_{j}\}$, thus we write $E_{n\sigma}(q)$ to emphasize this fact. So, the partition function reads
\begin{equation}
\mathcal{Z}=\sum_{\{q_{j}\}}\prod_{n\sigma}(1+e^{-\beta E_{n\sigma}(q)})=\sum_{\{q_{j}\}}e^{-\beta F(q)}\nonumber
\end{equation}
where we have defined an effective free energy
\begin{equation}
F(q)=-T\sum_{n\sigma}\ln(1+e^{-\beta E_{n\sigma}(q)}).\label{eq3}
\end{equation}
In this situation, we can explain $e^{-\beta F(q)}$ or $\rho(q)=\frac{1}{\mathcal{Z}}e^{-\beta F(q)}$ as an effective Boltzmann weight for each configuration of $\{q_{j}\}$ and this can be used to perform Monte Carlo simulation just like the classic Ising model.

To calculate physical quantities, we consider generic operator $\hat{O}$, which can be split into part with only Ising spin $\{q_{j}\}$ and another part with fermions,
\begin{equation}
\hat{O}=\hat{O}^{c}+\hat{O}^{q}.\nonumber
\end{equation}
Then, its expectation value in the equilibrium ensemble reads
\begin{equation}
\langle\hat{O}\rangle=\langle\hat{O}^{c}\rangle+\langle\hat{O}^{q}\rangle=\frac{\mathrm{Tr} \hat{O}^{c}e^{-\beta \hat{H}}}{\mathrm{Tr} e^{-\beta \hat{H}}}+\frac{\mathrm{Tr} \hat{O}^{q}e^{-\beta \hat{H}}}{\mathrm{Tr} e^{-\beta \hat{H}}}\nonumber
\end{equation}
For $\hat{O}^{q}$, we have
\begin{eqnarray}
\langle\hat{O}^{q}\rangle&&=\frac{\sum_{\{q_{j}\}}\hat{O}^{q}(q)e^{\beta h\sum_{j}q_{j}}\mathrm{Tr}_{c}e^{-\beta \hat{H}(q)}}{\sum_{\{q_{j}\}}e^{-\beta F(q)}}\nonumber\\
&&=\frac{\sum_{\{q_{j}\}}\hat{O}^{q}(q)e^{-\beta F(q)}}{\sum_{\{q_{j}\}}e^{-\beta F(q)}}\nonumber\\
&&=\sum_{\{q_{j}\}}\hat{O}^{q}(q)\rho(q)\nonumber.
\end{eqnarray}
In the Metropolis importance sampling algorithm, the above equation means we can use the simple average to estimate the expectation value like
\begin{equation}
\langle\hat{O}^{q}\rangle\simeq \frac{1}{N_{m}}\sum_{\{q_{j}\}} \hat{O}^{q}(q)\label{eq4}
\end{equation}
where $N_{m}$ is the number of sampling and the sum is over each configuration. $\hat{O}^{q}(q)$ is a number since we always work on the basis of $\{q_{j}\}$.

For $\hat{O}^{c}$,
\begin{equation}
\langle\hat{O}^{c}\rangle=\frac{\sum_{\{q_{j}\}}e^{\beta h\sum_{j}q_{j}}\mathrm{Tr}_{c}\hat{O}^{c}(q)e^{-\beta \hat{H}(q)}}{\sum_{\{q_{j}\}}e^{-\beta F(q)}}\nonumber
\end{equation}
and we can insert $\frac{e^{-\beta F(q)}}{e^{-\beta F(q)}}$ in the numerator, which leads to
\begin{eqnarray}
\langle\hat{O}^{c}\rangle&&=\sum_{\{q_{j}\}}\frac{\mathrm{Tr}_{c}\hat{O}^{c}(q)e^{-\beta \hat{H}(q)}}{\mathrm{Tr}_{c}e^{-\beta \hat{H}(q)}}\frac{e^{-\beta F(q)}}{\sum_{\{q_{j}\}}e^{-\beta F(q)}}\nonumber\\
&&=\sum_{\{q_{j}\}}\frac{\mathrm{Tr}_{c}\hat{O}^{c}(q)e^{-\beta \hat{H}(q)}}{\mathrm{Tr}_{c}e^{-\beta \hat{H}(q)}}\rho(q)\nonumber\\
&&=\sum_{\{q_{j}\}}\langle\langle\hat{O}^{c}(q)\rangle\rangle\rho(q)\nonumber.
\end{eqnarray}
This means
\begin{equation}
\langle\hat{O}^{c}\rangle\simeq \frac{1}{N_{m}}\sum_{\{q_{j}\}} \langle\langle\hat{O}^{c}(q)\rangle\rangle,\label{eq5}
\end{equation}
where $\langle\langle\hat{O}^{c}(q)\rangle\rangle=\frac{\mathrm{Tr}_{c}\hat{O}^{c}(q)e^{-\beta \hat{H}(q)}}{\mathrm{Tr}_{c}e^{-\beta \hat{H}(q)}}$ is calculated based on the Hamiltonian $\hat{H}(q)$. More practically, such statement means if fermions are involved, one can just calculate with $\hat{H}(q)$. Then, average over all sampled configuration gives rise to desirable results.

\section{Correlation function and spectral function}\label{apB}
When we calculate fermion's correlation function like $\langle \hat{c}_{i}^{\dag}\hat{c}_{j}\hat{c}_{k}^{\dag}\hat{c}_{l}\rangle$,
\begin{equation}
\langle \hat{c}_{i}^{\dag}\hat{c}_{j}\hat{c}_{k}^{\dag}\hat{c}_{l}\rangle=\frac{1}{N_{m}}\sum_{\{q_{j}\}}\langle\langle \hat{c}_{i}^{\dag}\hat{c}_{j}\hat{c}_{k}^{\dag}\hat{c}_{l}\rangle\rangle.\nonumber
\end{equation}
Then, using the Wick theorem for these free fermions, we get
\begin{eqnarray}
\langle\langle\hat{c}_{i}^{\dag}\hat{c}_{j}\hat{c}_{k}^{\dag}\hat{c}_{l}\rangle\rangle&&=\langle\langle\hat{c}_{i}^{\dag}\hat{c}_{j}\rangle\rangle\langle\langle\hat{c}_{k}^{\dag}\hat{c}_{l}\rangle\rangle+\langle\langle\hat{c}_{i}^{\dag}\hat{c}_{l}\rangle\rangle\langle\langle\hat{c}_{j}\hat{c}_{k}^{\dag}\rangle\rangle.\nonumber
\end{eqnarray}
Next, for each one-body correlation function like $\langle\langle\hat{c}_{i}^{\dag}\hat{c}_{j}\rangle\rangle$, one can transform these objects into their quasiparticle basis,
\begin{eqnarray}
g_{ij}^{q}\equiv\langle\langle \hat{c}_{i}^{\dag}\hat{c}_{j}\rangle\rangle&&=\sum_{m,n}\langle\langle\hat{d}_{m}^{\dag}\hat{d}_{n}\rangle\rangle(\phi_{m}^{i})^{\ast}\phi_{n}^{j}\nonumber\\
&&=\sum_{m,n}f_{F}(E_{n}(q))\delta_{m,n}(\phi_{m}^{i})^{\ast}\phi_{n}^{j}\nonumber\\
&&=\sum_{n}f_{F}(E_{n}(q))(\phi_{n}^{i})^{\ast}\phi_{n}^{j}\nonumber,
\end{eqnarray}
Similarly, we have
\begin{equation}
\langle\langle \hat{c}_{i}\hat{c}_{j}^{\dag}\rangle\rangle=\sum_{n}(1-f_{F}(E_{n}(q)))\phi_{n}^{i}(\phi_{n}^{j})^{\ast}=\delta_{ij}-g_{ji}^{q}.\nonumber
\end{equation}

For calculating dynamic quantities like conductance, (imaginary) time-dependent correlation function such as $\langle\langle \hat{c}_{i}^{\dag}(\tau)\hat{c}_{j}\rangle\rangle, \langle\langle \hat{c}_{i}(\tau)\hat{c}_{j}^{\dag}\rangle\rangle$ has to be considered. It is easy to show that,
\begin{eqnarray}
&&\hat{c}_{i}^{\dag}(\tau)=\sum_{n}\hat{d}_{n}^{\dag}(\tau)(\phi_{n}^{i})^{\ast}=\sum_{n}\hat{d}_{n}^{\dag}(\phi_{n}^{i})^{\ast}e^{\tau E_{n}(q)}\nonumber\\
&&\hat{c}_{i}(\tau)=\sum_{n}\hat{d}_{n}(\tau)\phi_{n}^{i}=\sum_{n}\hat{d}_{n}\phi_{n}^{i}e^{-\tau E_{n}(q)}.\nonumber
\end{eqnarray}
Therefore,
\begin{eqnarray}
&&\langle\langle \hat{c}_{i}^{\dag}(\tau)\hat{c}_{j}\rangle\rangle=\sum_{n}f_{F}(E_{n}(q))e^{\tau E_{n}(q)}(\phi_{n}^{i})^{\ast}\phi_{n}^{j}\nonumber\\
&&\langle\langle \hat{c}_{i}(\tau)\hat{c}_{j}^{\dag}\rangle\rangle=\sum_{n}(1-f_{F}(E_{n}(q)))e^{-\tau E_{n}(q)}\phi_{n}^{i}(\phi_{n}^{j})^{\ast}\nonumber\\
&&\langle\langle \hat{c}_{i}^{\dag}\hat{c}_{j}(\tau)\rangle\rangle=\sum_{n}f_{F}(E_{n}(q))e^{-\tau E_{n}(q)}(\phi_{n}^{i})^{\ast}\phi_{n}^{j}\nonumber\\
&&\langle\langle \hat{c}_{i}\hat{c}_{j}^{\dag}(\tau)\rangle\rangle=\sum_{n}(1-f_{F}(E_{n}(q)))e^{\tau E_{n}(q)}\phi_{n}^{i}(\phi_{n}^{j})^{\ast}.\nonumber
\end{eqnarray}
Using time-dependent correlation function, the imaginary-time Green's function for fixed Ising spin configuration is
\begin{eqnarray}
G_{ij}^{q}(\tau)&&=-\langle\langle T_{\tau}\hat{c}_{i}(\tau)\hat{c}_{j}^{\dag}\rangle\rangle\nonumber\\
&&=-\theta(\tau)\langle\langle\hat{c}_{i}(\tau)\hat{c}_{j}^{\dag}\rangle\rangle+\theta(-\tau)\langle\langle\hat{c}_{j}^{\dag}\hat{c}_{i}(\tau)\rangle\rangle\nonumber.
\end{eqnarray}
Thus, if assuming $\tau>0$, one finds the following relations between time-dependent correlation functions and their Green's function,
\begin{eqnarray}
&&G_{ji}^{q}(-\tau)=\langle\langle \hat{c}_{i}^{\dag}(\tau)\hat{c}_{j}\rangle\rangle,G_{ij}^{q}(\tau)=-\langle\langle \hat{c}_{i}(\tau)\hat{c}_{j}^{\dag}\rangle\rangle\nonumber\\
&&G_{ji}^{q}(\tau)=\langle\langle \hat{c}_{i}^{\dag}\hat{c}_{j}(\tau)\rangle\rangle,G_{ij}^{q}(-\tau)=-\langle\langle \hat{c}_{i}\hat{c}_{j}^{\dag}(\tau)\rangle\rangle\nonumber.
\end{eqnarray}
The Fourier transformation of $G_{ij}^{q}(\tau)$ reads
\begin{eqnarray}
G_{ij}^{q}(\omega_{n})=\int_{0}^{\beta}d\tau e^{i\omega_{n}\tau}G_{ij}^{q}(\tau)=\sum_{n}\frac{\phi_{n}^{i}(\phi_{n}^{j})^{\ast}}{i\omega_{n}-E_{n}(q)}.\nonumber
\end{eqnarray}
The corresponding retarded Green's function is obtained via analytic continuity $i\omega_{n}\rightarrow \omega+i0^{+}$
\begin{eqnarray}
G_{ij}^{q}(\omega)=\sum_{n}\frac{\phi_{n}^{i}(\phi_{n}^{j})^{\ast}}{\omega+i0^{+}-E_{n}(q)}.\nonumber
\end{eqnarray}
And the related spectral function is
\begin{eqnarray}
A_{ij}^{q}(\omega)=-\frac{1}{\pi}\mathrm{Im}G_{ij}^{q}(\omega)=\sum_{n}\phi_{n}^{i}(\phi_{n}^{j})^{\ast}\delta(\omega-E_{n}(q)).\nonumber
\end{eqnarray}
The spectral function with momentum-dependence has essential importance to spectral experiments, which can be found as
\begin{eqnarray}
A^{q}(k,\omega)=\frac{1}{N_{s}}\sum_{ij}e^{-ik\cdot(R_{i}-R_{j})}A_{ij}^{q}(\omega).\nonumber
\end{eqnarray}
\section{Static conductance and resistance}\label{apC}
The dc conductance is related to current-current correlation function as
\begin{equation}
\sigma_{dc}=\lim_{\omega\rightarrow0}\frac{\mathrm{Im} \Lambda_{xx}(k=0,\omega)}{\omega}\nonumber
\end{equation}
and the static resistivity is $\rho=1/\sigma_{dc}$. Here, the retarded current-current correlation function $\Lambda_{xx}(k=0,\omega)$ can be obtained via its imaginary-time form
\begin{equation}
\Lambda_{xx}(k,i\Omega_{n})=\frac{1}{N_{s}}\sum_{i,j}e^{ik(R_{i}-R_{j})}\int d\tau e^{i\Omega_{n}\tau}\langle \hat{J}_{x}(i,\tau)\hat{J}_{x}(j,0)\rangle.\nonumber
\end{equation}
Here, $\hat{J}_{x}$ is $x$-axis component of current operator.
Because our model is defined on a lattice, in terms of Peierls substitution, the external electromagnetic potential $A_{x}(i)\equiv A_{i,i+x}$ is introduced as
\begin{equation}
-t\sum_{\langle ij\rangle,\sigma}\hat{c}_{i\sigma}^{\dag}\hat{c}_{j\sigma}\rightarrow-t\sum_{\langle ij\rangle,\sigma}e^{ieA_{ij}}\hat{c}_{i\sigma}^{\dag}\hat{c}_{j\sigma}\nonumber
\end{equation}
with $A_{ij}=-A_{ji}$. Now, $\hat{J}_{x}(i)$ is derived as
\begin{equation}
\hat{J}_{x}(i)=\lim_{A\rightarrow0}\frac{\delta \hat{H}}{\delta A_{x}(i)}=ite\sum_{\sigma}(\hat{c}_{i+x,\sigma}^{\dag}\hat{c}_{i\sigma}-\hat{c}_{i\sigma}^{\dag}\hat{c}_{i+x,\sigma}).\nonumber
\end{equation}
Therefore,
\begin{equation}
\langle\hat{J}_{x}(i,\tau)\hat{J}_{x}(j,0)\rangle=\frac{1}{N_{m}}\sum_{\{q\}}\langle\langle\hat{J}_{x}(i,\tau)\hat{J}_{x}(j,0)\rangle\rangle\nonumber
\end{equation}
and
\begin{eqnarray}
\frac{1}{(ite)^{2}}\langle\langle\hat{J}_{x}(i,\tau)\hat{J}_{x}(j,0)\rangle\rangle&&=\sum_{ij,\sigma\sigma'}g_{i,i+x,\sigma}^{q}(g_{j,j+x,\sigma'}^{q}-g_{j+x,j,\sigma'}^{q})\nonumber\\
&&+g_{i+x,i,\sigma}^{q}(g_{j+x,j,\sigma'}^{q}-g_{j,j+x,\sigma'}^{q})\nonumber\\
&&-\delta_{\sigma\sigma'}G_{j+x,i,\sigma}^{q}(-\tau)G_{i+x,j,\sigma}^{q}(\tau)\nonumber\\
&&+\delta_{\sigma\sigma'}G_{j,i,\sigma}^{q}(-\tau)G_{i+x,j+x,\sigma}^{q}(\tau)\nonumber\\
&&+\delta_{\sigma\sigma'}G_{j+x,i+x,\sigma}^{q}(-\tau)G_{i,j,\sigma}^{q}(\tau)\nonumber\\
&&-\delta_{\sigma\sigma'}G_{j,i+x,\sigma}^{q}(-\tau)G_{i,j+x,\sigma}^{q}(\tau)\nonumber.
\end{eqnarray}
Here, $g_{ij\sigma}^{q}$ has no frequency-dependence and imaginary part, thus it cannot contribute to conductance and will be neglected hereafter.
Integrating over $\tau$ gives
\begin{eqnarray}
&&\int d\tau e^{i\Omega_{n}\tau}\frac{1}{(ite)^{2}}\langle\langle \hat{J}_{x}(i,\tau)\hat{J}_{x}(j,0)\rangle\rangle\nonumber\\
=&&-T\sum_{\omega_{n},\sigma}G_{j+x,i,\sigma}^{q}(\omega_{n})G_{i+x,j,\sigma}^{q}(\omega_{n}+\Omega_{n})\nonumber\\
&&+T\sum_{\omega_{n},\sigma}G_{j,i,\sigma}^{q}(\omega_{n})G_{i+x,j+x,\sigma}^{q}(\omega_{n}+\Omega_{n})\nonumber\\
&&+T\sum_{\omega_{n},\sigma}G_{j+x,i+x,\sigma}^{q}(\omega_{n})G_{i,j,\sigma}^{q}(\omega_{n}+\Omega_{n})\nonumber\\
&&-T\sum_{\omega_{n},\sigma}G_{j,i+x,\sigma}^{q}(\omega_{n})G_{i,j+x,\sigma}^{q}(\omega_{n}+\Omega_{n})\nonumber\\
=&&\sum_{\sigma}\int d\omega_{1}\int d\omega_{2}\frac{f_{F}(\omega_{1})-f_{F}(\omega_{2})}{i\Omega_{n}-\omega_{2}+\omega_{1}}
\nonumber\\
&&\times[
-A_{j+x,i,\sigma}^{q}(\omega_{1})A_{i+x,j,\sigma}^{q}(\omega_{2})\nonumber\\
&&+A_{j,i,\sigma}^{q}(\omega_{1})A_{i+x,j+x,\sigma}^{q}(\omega_{2})
+A_{j+x,i+x,\sigma}^{q}(\omega_{1})A_{i,j,\sigma}^{q}(\omega_{2})\nonumber\\
&&-A_{j,i+x,\sigma}^{q}(\omega_{1})A_{i,j+x,\sigma}^{q}(\omega_{2})],\nonumber
\end{eqnarray}
which leads to the retarded current-current correlation
\begin{eqnarray}
&&\Lambda_{xx}(k,\omega+i0^{+})=\frac{(ite)^{2}}{N_{s}N_{m}}\sum_{\{q\}}\sum_{ij,\sigma}\int d\omega_{1}\int d\omega_{2}\nonumber\\
&&\frac{f_{F}(\omega_{1})-f_{F}(\omega_{2})}{\omega+i0^{+}-\omega_{2}+\omega_{1}}e^{ik(R_{i}-R_{j})}\nonumber\\
&&\times[
-A_{j+x,i,\sigma}^{q}(\omega_{1})A_{i+x,j,\sigma}^{q}(\omega_{2})+A_{j,i,\sigma}^{q}(\omega_{1})A_{i+x,j+x,\sigma}^{q}(\omega_{2})\nonumber\\
&&+A_{j+x,i+x,\sigma}^{q}(\omega_{1})A_{i,j,\sigma}^{q}(\omega_{2})-A_{j,i+x,\sigma}^{q}(\omega_{1})A_{i,j+x,\sigma}^{q}(\omega_{2})].\nonumber
\end{eqnarray}
Now, it is straightforward to get
\begin{eqnarray}
&&\mathrm{Im}\Lambda_{xx}(0,\omega)=\frac{\pi t^{2}e^{2}}{N_{s}N_{m}}\sum_{\{q\}}\sum_{ij,\sigma}\int d\omega_{1}\left(f_{F}(\omega_{1})-f_{F}(\omega_{1}+\omega)\right)\nonumber\\
&&\times[
-A_{j+x,i,\sigma}^{q}(\omega_{1})A_{i+x,j,\sigma}^{q}(\omega_{1}+\omega)\nonumber\\
&&+A_{j,i,\sigma}^{q}(\omega_{1})A_{i+x,j+x,\sigma}^{q}(\omega_{1}+\omega)\nonumber\\
&&+A_{j+x,i+x,\sigma}^{q}(\omega_{1})A_{i,j,\sigma}^{q}(\omega_{1}+\omega)\nonumber\\
&&-A_{j,i+x,\sigma}^{q}(\omega_{1})A_{i,j+x,\sigma}^{q}(\omega_{1}+\omega)].\nonumber
\end{eqnarray}
Finally, the dc conductance is found to be
\begin{eqnarray}
\sigma_{dc}=\frac{\pi t^{2}e^{2}}{N_{m}}\sum_{\{q\}}\int d\omega\frac{-\partial f_{F}(\omega)}{\partial \omega}\Phi^{q}(\omega)\nonumber
\end{eqnarray}
with
\begin{eqnarray}
\Phi^{q}(\omega)&&=\frac{1}{N_{s}}\sum_{ij,\sigma}[-A_{j+x,i,\sigma}^{q}(\omega)A_{i+x,j,\sigma}^{q}(\omega)+A_{j,i,\sigma}^{q}(\omega)A_{i+x,j+x,\sigma}^{q}(\omega)\nonumber\\
&&+A_{j+x,i+x,\sigma}^{q}(\omega)A_{i,j,\sigma}^{q}(\omega)-A_{j,i+x,\sigma}^{q}(\omega)A_{i,j+x,\sigma}^{q}(\omega)].\nonumber
\end{eqnarray}

\section{Example for doped system at $T=\infty$}\label{apD}
Here, we show the entanglement entropy $S_{EE}$ and IPR at Fermi energy $\mathrm{IPR(0)}$ for the doped system. We choose chemical potential $\mu$ as
$\mu/t=-4,-3,-2,-1,0$ and set $J/t=8$. Because the low $T$ phase diagram of the doped system is rather complicated due to intertwined magnetic orders, instead, we focus on $T=\infty$ limit, where only paramagnetic phases survive.
\begin{figure}
\includegraphics[width=1.1\linewidth]{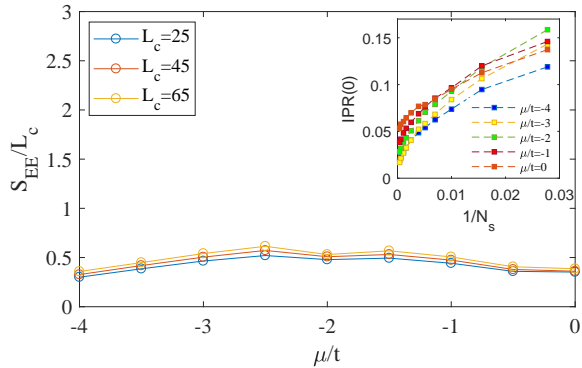}
\caption{\label{fig:7} $S_{EE}$ and $\mathrm{IPR(0)}$ versus chemical potential $\mu$ for the doped system with $J/t=8$ at $T=\infty$.}
\end{figure}

Using Eq.~\ref{eq11}, $S_{EE}$ and $\mathrm{IPR(0)}$ are shown in Fig.~\ref{fig:7}. We find that $S_{EE}$ for different $\mu$ has similar linear dependence on $L_{c}$, and $\mathrm{IPR(0)}$ in infinite system limit is finite. Thus, the AL phase is stable when deviating from half-filling, at least in $T=\infty$ limit.
\section{Finite IPR at finite $T$}\label{apE}

\begin{figure}[htb]
\includegraphics[width=1.05\linewidth]{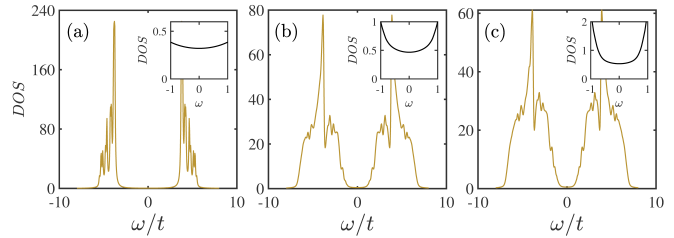}
\caption{\label{fig:apDOS} DOS of conduction electron $N(\omega)$ in MI (J/t=15) at different temperature (a) $T/t=0.1$, (b) $T/t=0.4$, (c) $T/t=0.8$. With increasing temperature, the DOS at Fermi surface increases and the gap decreases.}
\end{figure}
\begin{figure}[htb]
\includegraphics[width=0.63\linewidth]{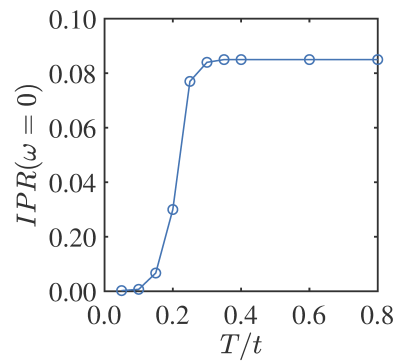}
\caption{\label{fig:apIPR} The IPR versus temperature at $J/t=15$, which is calculated at thermodynamic limit.}
\end{figure}

At zero-temperature, due to the absence of eigenstates in the gap of Mott insulator, the IPR at Fermi surface should be strictly equal to zero. However, at finite temperature situation, the physics of excited states also contributes to the IKL system, leading to finite DOS (see Fig.~\ref{fig:apDOS}) and IPR (see Fig.~\ref{fig:apIPR}) in the Mott gap. As shown in Fig.~\ref{fig:apIPR}, with decreasing temperature the IPR$(\omega=0)$ monotonically decreases, which is approaching to zero around $T=0$. Similar results about nonzero IPR$(\omega=0)$ at finite temperature could also be found in previous studies \cite{Antipov}.


\begin{thebibliography}{58}%
\bibitem{Anderson} Anderson P W 1958 \textit{Phys. Rev. B}  \textbf{109} 1492
\bibitem{Lee} Lee P A and Ramakrishnan T V 1985 \textit{Rev. Mod. Phys.} \textbf{57} 287
\bibitem{Evers} Evers F and Mirlin A D 2008 \textit{Rev. Mod. Phys.} \textbf{80} 1355
\bibitem{Gornyi} Gornyi I V, Mirlin A D and Polyakov D G 2005 \textit{Phys. Rev. Lett.} \textbf{95} 206603
\bibitem{Basko} Basko D M, Aleiner I L and Altshuler B L 2006 \textit{Ann. Phys.}  \textbf{321} 1126
\bibitem{Nandkishore} Nandkishore R and Huse D A 2015 \textit{Annu. Rev. Condens. Matter Phys.} \textbf{6} 15
\bibitem{Abanin}  Abanin D A, Altman E, Bloch I and Serbyn M 2019 \textit{Rev. Mod. Phys.} \textbf{91} 021001
\bibitem{Antipov} Antipov A E, Javanmard Y, Ribeiro P and Kirchner S 2016 \textit{Phys. Rev. Lett.} \textbf{117} 146601
\bibitem{Castro} Gon\c{c}alves M, Ribeiro P, Mondaini R and Castro E V 2019 \textit{Phys. Rev. Lett.} \textbf{122} 126601
\bibitem{Smith} Smith A, Knolle J, Kovrizhin D L and Moessner R 2017 \textit{Phys. Rev. Lett.} \textbf{118} 266601
\bibitem{Smith2} Smith A, Knolle J, Moessner R and Kovrizhin D L 2018 \textit{Phys. Rev. B} \textbf{97} 245137
\bibitem{Srednicki} Srednicki M 1994 \textit{Phys. Rev. E} \textbf{50} 888
\bibitem{Rigol} Rigol M, Dunjko V and Olshanii M 2008 \textit{Nature} \textbf{452} 854
\bibitem{Kogut} Kogut J B 1979 \textit{Rev. Mod. Phys.} \textbf{51} 659
\bibitem{Smith3} Smith A, Knolle J, Moessner R and Kovrizhin D L 2017 \textit{Phys. Rev. Lett.} \textbf{119} 176601
\bibitem{Sikkema} Sikkema A E, Buyers W J L, Affleck I and Gan J 1996 \textit{Phys. Rev. B } \textbf{54} 9322
\bibitem{Yang} Yang W W, Zhao J Z, Luo H G and Zhong Y 2019 \textit{Phys. Rev. B} \textbf{100} 045148
\bibitem{White} White S R and Scalapino D J 1998 \textit{Phys. Rev. Lett.} \textbf{80} 1272
\bibitem{Lynn} Lynn J W, Skanthakumar S, Huang Q, Sinha S K, Hossain Z, Gupta L C,
Nagarajan R and Godart C 1997 \textit{Phys. Rev. B} \textbf{55} 6584
\bibitem{Falicov} Falicov L M and Kimball J C 1969 \textit{Phys. Rev. Lett.} \textbf{22} 997
\bibitem{Mydosh} Mydosh J A, Oppeneer P M 2011 \textit{Rev. Mod. Phys.} \textbf{83} 1301
\bibitem{Coleman2015} Coleman P 2015 \textit{Introduction to Many Body Physics} (Cambridge: Cambridge University Press) pp.486-580
\bibitem{Si} Si Q and Paschen S 2013 \textit{Phys. Stat. Solid. B} \textbf{250} 425-438
\bibitem{Coleman} Coleman P and Nevidomskyy A H 2010 \textit{J. Low Temp. Phys.} \textbf{161} 182
\bibitem{Kitaev1}  Kitaev A 2003 \textit{Ann. Phys.} \textbf{303} 2
\bibitem{Kitaev2} Kitaev A 2006 \textit{Ann. Phys.} \textbf{321} 2
\bibitem{Czajka} Maska M M and Czajka K 2006 \textit{Phys. Rev. B} \textbf{74} 035109
\bibitem{Yunoki} Yunoki S, Hu J, Malvezzi A L, Moreo A, Furukawa N and Dagotto E 1998 \textit{Phys. Rev. Lett.} \textbf{80} 845
\bibitem{Dagotto} Dagotto E, Yunoki S, Malvezzi A L, Moreo A, Hu J, Capponi S, Poilblanc D and Furukawa N 1998 \textit{Phys. Rev. B} \textbf{58} 6414
\bibitem{Thoss} \u{Z}onda M, Okamoto J and Thoss M 2019 \textit{Phys. Rev. B} \textbf{100} 075124
\bibitem{PhysRevB.90.024202} Malmi-Kakkada A N, Valls O T, and Dasgupta C 2014 \textit{Phys. Rev. B} \textbf{90} 024202
\bibitem{RevModPhys.58.801} Binder K and Young A P 1986 \textit{Rev. Mod. Phys.} \textbf{58} 801
\bibitem{PhysRevB.76.245122} Tran M T 2007 \textit{Phys. Rev. B} \textbf{76} 245122
\bibitem{Byczuk} Byczuk K, Hofstetter W and Vollhardt D 2010 \textit{International Journal of Modern Physics B} \textbf{24} 1727
\bibitem{Dzero} Dzero M, Sun K, Coleman P, and Galitski V 2012 \textit{Phys. Rev. B} \textbf{85} 045130
\bibitem{Eisert} Eisert J, Cramer M and Plenio M B 2010 \textit{Rev. Mod. Phys.} \textbf{82} 277
\bibitem{Wolf} Wolf M M 2006 \textit{Phys. Rev. Lett.} \textbf{96} 010404
\bibitem{Gioev} Gioev D and Klich I 2006 \textit{Phys. Rev. Lett.} \textbf{96} 100503
\bibitem{Hubbard} Hubbard J 1959 \textit{Phys. Rev. Lett.} \textbf{3} 77
\bibitem{Stratonovich} Stratonovich R L 1958 \textit{Sov. Phys.-Doklady} \textbf{2}
\end{thebibliography}
\end{document}